\documentclass[prd,preprint,eqsecnum,nofootinbib,amsmath,amssymb,
               tightenlines,dvips]{revtex4}
\usepackage{graphicx}
\usepackage{bm}%	bold math

\begin {document}

%%%%%%%%%%%%%%%%%%%%%%%%%%%%%%%%%%%%%%%%%%%%%%%%%%%%%%%%%%%%%%%%%%%%%%%%%%%%%%%

\def\Mrowczynski{Mr\'owczy\'nski}

\def\D{{\bm D}}

\def\k{{\bm k}}
\def\half{{\textstyle{\frac12}}}

\def\p{{\bm p}}

\def\x{{\bm x}}

\def\v{{\bm v}}
\def\E{{\bm E}}
\def\B{{\bm B}}
\def\A{{\bm A}}

\def\grad{{\bm\nabla}}

\def\ms{m_{\rm s}}

%%%%%%%%%%%%%%%%%%%%%%%%%%%%%%%%%%%%%%%%%%%%%%%%%%%%%%%%%%%%%%%%%%%%%%%%%%%%%%%

\title
    {
    Lessons From Non-Abelian Plasma Instabilities in Two Spatial Dimensions    
    }

\author{Peter Arnold and Po-Shan Leang}
 \affiliation
    {%
    Department of Physics,
    University of Virginia, P.O. Box 400714
    Charlottesville, Virginia 22904-4714, USA
    }%

\date {April, 2007}

\begin {abstract}%
    {%
       Plasma instabilities can play a fundamental role in quark-gluon
       plasma equilibration in the high energy (weak coupling) limit.
       Early simulations of the evolution of plasma instabilities in
       non-abelian gauge
       theory, performed in one spatial dimension, found
       behavior qualitatively similar to traditional QED plasmas.
       Later simulations of the fully three-dimensional theory found
       different behavior, unlike traditional QED plasmas.
       To shed light on the origin of
       this difference, we study the intermediate case of two
       spatial dimensions.  Depending on how the
       ``two-dimensional'' theory is formulated, we can obtain either
       behavior.
    }
\end {abstract}

\maketitle
\thispagestyle {empty}

%%%%%%%%%%%%%%%%%%%%%%%%%%%%%%%%%%%%%%%%%%%%%%%%%%%%%%%%%%%%%%%%%%%%%%%%%%%%%%%

\section {Introduction and Results}
\label{sec:intro}

It is a difficult theoretical problem to fully understand how
quark-gluon plasmas come to local equilibrium in the context of heavy
ion collisions.  In fact, equilibration is not yet fully understood even
in the formal limit of weakly-coupled quark-gluon plasmas
(equivalently the limit of
arbitrarily high energy collisions).  By investigating the rates and
interplay of individual scattering processes, Baier, Mueller, Schiff and
Son \cite {bottom_up} attempted to analyze equilibration in the weak
coupling limit, producing what is known as the bottom-up scenario
of quark-gluon plasma equilibration.
However, interesting collective behavior in the form of
plasma instabilities
turns out to severely complicate the problem
\cite{ALM}.
The relevant instabilities are known in traditional plasma physics as
Weibel or filamentary instabilities
\cite{weibel,plasma_old,RS}.
Over the last few years, there has been a variety of work
on the dynamics of plasma
instabilities in the weakly-coupled limit of non-abelian gauge theory
\cite{AL,RRS,linear1,RRS2,Nara,linear2,DNS,RV,kminus2,MSWnewBUP,BnewBUP,MSW}.
Plasma instabilities initially grow exponentially quickly, and they
create large color magnetic fields which could speed up local
isotropization and thermalization of the initially non-equilibrium
plasma by scattering
plasma particles into random directions.  To puzzle together the theory
of equilibration in the weak coupling limit, it is
important to understand just how large these instabilities and
their associated magnetic fields grow.  It has been of particular
interest to understand the similarities and differences between
the evolution of instabilities in non-abelian gauge theory plasmas
vs.\ traditional abelian plasmas \cite{linear1,RRS2,linear2,DNS}.

\begin{figure}[ht]
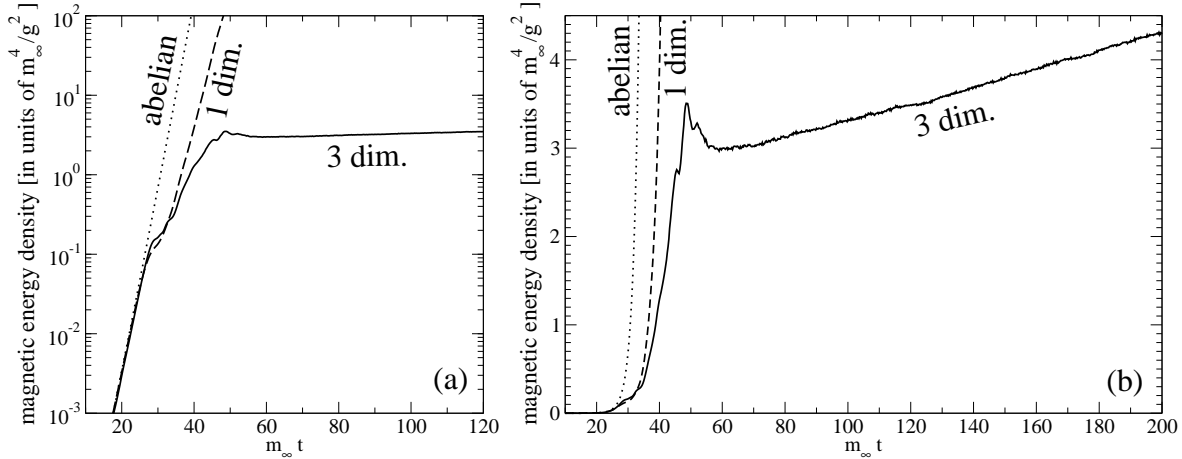

\includegraphics[scale=0.35]{linear1.eps}
\includegraphics[scale=0.35]{linear1_linear.eps}
\caption{%
    \label{fig:linear1}
    The magnetic field energy associated with instabilities as a
    function of time for abelian theories (dotted line),
    non-abelian theories restricted to one spatial dimensions (dashed line),
    and non-abelian theories in three spatial dimensions (solid line).
    The vertical axis is shown as both (a) logarithmic, and (b) linear.
    These particular simulations are taken from Ref.\ \cite{linear1}, but
    similar one-dimensional simulations were performed earlier in
    Ref.\ \cite{RRS}.  The non-Abelian simulations are based on SU(2) gauge
    theory.
}
\end{figure}

As we'll briefly review,
simulations have been previously
devised to test whether non-abelian effects
would limit the growth of plasma instabilities compared to 
abelian plasmas.  An example of the results \cite{linear1,RRS2} is shown in
Fig.\ \ref{fig:linear1}.  This figure shows the growth in time
of the magnetic energy of long-wavelength fields associated
with the instability.  In these simulations, the instability has
been seeded with a very small initial amplitude.
The
dotted line shows the abelian case, corresponding to simple
exponential growth of the instability (with a rate that can
be computed in perturbation theory).  This energy growth is
fueled by stealing energy from the particles in the plasma,
and it must eventually stop once the fraction of energy stolen
becomes significant.  In these simulations, that scale is
beyond the top of the graph.
(In fact, the back-reaction
on the plasma particles is intentionally ignored in these
simulations, providing an unlimited source of energy,
so that one
may cleanly disentangle whether or not non-abelian effects can limit
the instability growth earlier.)

For computational simplicity, the first simulations of the non-abelian
theory
were restricted to one spatial dimension \cite{RRS}, in a specific
sense reviewed later.
{}From rough reasoning based on the one-dimensional theory,
Ref.\ \cite{AL} had earlier conjectured that non-abelian interactions
would not stop exponential growth of the instability.
This conjecture was apparently supported by the one-dimensional
simulations, an example of which is shown by the dashed line in
Fig.\ \ref{fig:linear1}.  The growth starts out looking like the abelian case
(which is expected from perturbative analysis of the instability).
Eventually, there is a deviation when the fields get large enough
for non-linear self-interactions in the non-abelian theory to become
important, but the subsequently growth is again exponential.
The conjecture proved wrong for the full three-dimensional theory,
however, shown by the solid line in Fig.\ \ref{fig:linear1}.
After non-linear interactions become important, there is a spurt
of rapid growth similar to the one-dimensional case,
but then instability growth stops
in three dimensions.
There is continued linear growth in the magnetic energy at late
times, shown in the linear plot of Fig.\ \ref{fig:linear1}b,
but this is due to the
unstable modes pumping energy into slightly higher momentum, stable
modes rather than by
continued growth of the unstable modes \cite{linear2}.

The purpose of this paper is to investigate the root cause of the
different behavior in one and three dimensions.
Our tool will be
the intermediate case of two dimensions.
Before explaining our results,
however, we need to review the three-dimensional
theory being simulated and how one defines its reduction
to lower dimensions.
Following the review, we
will present our two-dimensional results.
We can get either one-dimensional or three-dimensional behavior,
depending on how the two-dimensional theory is constructed.
Then, in Sec.\ \ref{sec:interpret}, we offer a general interpretation of
our results in terms of whether all polarizations of the gauge field
are gauged in the dimensionally-reduced theory.
For the sake of concreteness, we also discuss a specific
example of this interpretation in the
context of Nielsen-Olesen instabilities, which could possibly play
a role in the fate of non-abelian plasma instabilities.
Finally, we offer our conclusions in Sec.\ \ref{sec:conclusion}.
Technical details of our simulations are relegated to an appendix.

% -------------------------------------------------------------------------

\subsection {Review}

Weibel instabilities occur when the momentum distribution of a
non-equilibrium plasma is anisotropic, the expansion and collision
rates are small compared to the instability growth rate, and
there is a clear separation of scale between the momentum of typical
plasma particles and the wave numbers associated with the resulting
instability.
The problem can be studied by simulating a non-abelian analog of
the classical Vlasov equations of traditional plasma physics.
In the weak coupling limit, one can justify the preceding assumptions
and the application of these classical equations to the 
plasma instabilities in
the bottom-up picture \cite{ALM}.
In the bottom-up picture, the energy stored in the plasma
particles is parametrically larger than the energy associated
with the onset of significant self-interactions of magnetic
fields associated with the instability.
The Vlasov equations consist of (i) a collisionless
Boltzmann equation describing the evolution of the phase space
density $f(\x,\p,t)$ of hard quarks and gluons in the presence of
a soft gauge field $A(\x,t)$, and (ii) Maxwell's equation for the
soft gauge field with sources given by the hard particles.
In order
to study how large the instabilities grow, it is sufficient to
linearize these equations in fluctuations $\delta f(\x,\p,t)$ of
the hard particles about an initial (homogeneous) distribution $f_0(\p)$,
while retaining the non-linear structure of the non-abelian gauge
field $A$.
This approximation treats the plasma particles as an infinite
reservoir of energy for instability growth.
Schematically (suppressing sums over particles types),
the equations are of the form%
\footnote{
  We work in $({-}{+}{+}{+})$ metric convention.
}
\begin {subequations}
\label {eq:basic}
\begin {equation}
    (D_t + \v \cdot \D_\x) \, \delta f
    +
    g \, (\E + \v \times \B) \cdot \grad_{\p} f_0
    =
    0 \, ,
\end {equation}
\begin {equation}
    D_\nu \, F^{\mu\nu} = j^\mu
    =
    g \nu t_R \int \frac{d^3p}{(2\pi)^3} \>  v^\mu \, \delta f \, .
\label {eq:basic1b}
\end {equation}
\end {subequations}
Above, $\delta f$ is in the adjoint representation, $D$ is the
corresponding covariant derivative, $t_R$ is a group factor,
$\nu$ counts particle spins and types,%
\footnote{
   For example, for a plasma of gluons, $\nu = 2$ and $t_R = 3$.
}
and $v^\mu
\equiv(1,\hat\p)$.  For ultra-relativistic systems, one can put these
equations in a more convenient form by integrating
over the magnitude $|\p|$ of momentum to replace $\delta f(\x,\p)$,
which depends on six phase-space coordinates, by a distribution
$W(\x,\v)$, which depends on only five since $\v=\hat\p$ lives
on the unit sphere.  $W^a(\x,\v)$ represents the net adjoint color
of all particles moving in direction $\v$ at point $\x$.
The corresponding Vlasov equations are
\begin {subequations}
\label {eq:basicW}
\begin {equation}
    (D_t + \v \cdot \D_\x) \, W
    + m_\infty^2 [\E\cdot(\grad_\v-2\v) - \B\cdot(\v\times\grad_\v)] \Omega(\v)
    =
    0 ,
\end {equation}
\begin {equation}
    D_\nu \, F^{\mu\nu} = j^\mu
    =
    \int_\v v^\mu W \, ,
\end {equation}
\end {subequations}
where $\Omega(\v)$ is determined by the angular dependence of the initial
particle distribution $f_0(\p)$, normalized so that the angular average
of $\Omega(\v)$ is one.
The mass scale $m_\infty$ above is set by the amplitude of $f_0$ as
\begin {equation}
   m_\infty^2 \equiv
   g^2 \nu t_R \int_0^\infty \frac{d^3p}{(2\pi)^3} \>
   \frac{f_0(p\v)}{p} \,.
\end {equation}
For moderately anisotropic plasmas, this is the order of magnitude of
instability wave-numbers, instability growth rates, plasmon masses, and
Debye screening, all of which we refer to as soft physics.
Soft magnetic fields become non-perturbative when $B \sim m_\infty^2/g$,
with corresponding energy density $\half B^2 \sim m_\infty^4/g^2$.

In this paper, we focus on the case of moderately anisotropic, oblate
velocity distributions which are axi-symmetric about the $z$ axis.
In particular, we will study precisely the same angular
distribution $\Omega(\v)$ studied in Ref.\ \cite{linear1}.
A full study of the bottom-up scenario will require simulating
extremely anisotropic distributions \cite{ALM,BnewBUP,kminus2},
which is computationally
more difficult.  However, moderately anisotropic distributions will
be adequate for the goal of this paper to understand the origin of
the differences between one and three dimensional simulations.
For the sake of numerical simplicity, our simulations are all based on SU(2)
gauge theory rather than SU(3) QCD.  We do not have any reason to
expect qualitative differences between the two.

The ``one dimensional'' version of equations
(\ref{eq:basic}) and (\ref{eq:basicW}) consists of
assuming that fields and distributions vary in space only in
one direction $z$, as $A_\mu(z,t)$ and $\delta f(z,\p,t)$ [or
$W(z,\v,t)$].  Particle momenta $\p$, however, are still
treated three-dimensionally, and all three spatial polarizations of the
gauge field ($A_x, A_y, A_z$) are included.  This type of dimensional
reduction of the full three dimensional theory is known in traditional
plasma physics as a 1D+3V theory, where the 1D signifies that $\x$ has
been reduced to one dimension, but the 3V indicates that velocity
(momentum) is still treated three-dimensionally.
Fully three-dimensional simulations are 3D+3V.
Note that if the transverse polarizations $A_x(z,t)$ and $A_y(z,t)$
were not included in one-dimensional simulations, one could not describe
magnetic physics at all and so could not investigate Weibel
instabilities.

% -------------------------------------------------------------------------

\subsection {Two-Dimensional Results}

\begin{figure}[ht]
\includegraphics[scale=0.60]{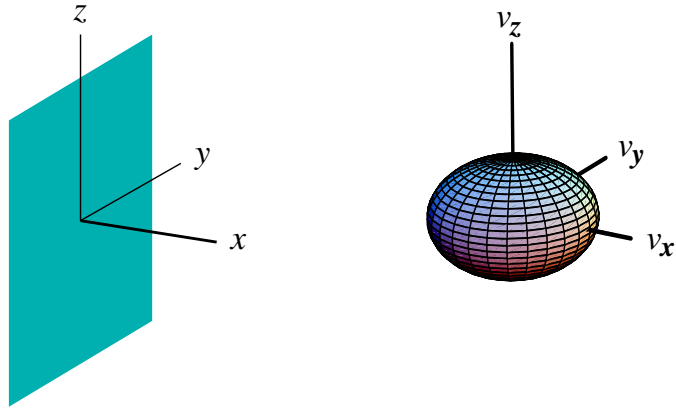}
\caption{%
    \label{fig:plane}
    Axis labeling for two-dimensional (2D+3V) simulations.
    The spatial plane $\x$ is the $yz$ plane, shown on the left,
    but particle momenta $\p$ live in three dimensions, and
    the initial distribution $\Omega(\v)$ of particle velocities
    is axi-symmetric about the $z$ axis, as depicted on the right.
}
\end{figure}

Fig.\ \ref{fig:results2d} shows our results for two-dimensional (2D+3V)
simulations.
The solid line represents the same kind of dimensional reduction that
was done in the one-dimensional case: all polarizations are retained,
but fields and distributions depend on only two spatial dimensions:%
\footnote{
  Though we have listed $\mu=t$ in (\ref{eq:2dA}) for the sake
  of generality, the simulations are carried out in
  $A_0=0$ gauge.  We will only discuss
  gauge-invariant results.
}
\begin {subequations}
\label {eq:2d}
\begin {align}
   A_\mu &= A_\mu(y,z,t), \qquad \mu=x,y,z,t ; \label{eq:2dA} \\
   \delta f &= \delta f(y,z,\p,t) ,
\end {align}
or $W = W(y,z,\v,t)$.
\end {subequations}
[The labeling of directions with respect to the two-dimensional plane
is shown in Fig.\ \ref{fig:plane} for reference.]
In this case, we find behavior similar to one-dimensional simulations:
instability growth continues exponentially even after non-abelian
interactions become important.
However, the dashed line shows what happens if we eliminate the
polarization $A_x$ which lies outside of the two-dimensional plane,
by setting $A_x=0$ in our two-dimensional simulations.
We now obtain behavior similar to three-dimensional
simulations instead: the energy growth at late times is linear rather than
exponential.

\begin{figure}[ht]
\includegraphics[scale=0.35]{results2d.eps}
\includegraphics[scale=0.35]{results2d_linear.eps}
\caption{%
    \label{fig:results2d}
    As Fig.\ \ref{fig:linear1} but for two-dimensional (2D+3V)
    simulations.  The solid line is for the straightforward
    dimensional reduction of (\ref{eq:2d}).  The dashed line
    is a simulation that does not include the out-of-plane
    polarization $A_x$.
}
\end{figure}

\begin{figure}[ht]
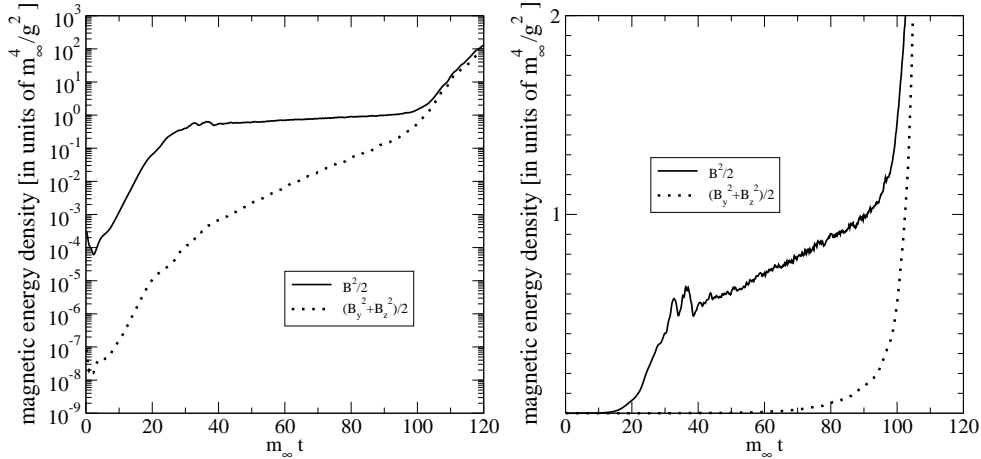

\includegraphics[scale=0.35]{results2dsmallmix.eps}
\includegraphics[scale=0.35]{results2dsmallmix_linear.eps}
\caption{%
    \label{fig:results2dsmallmix}
    As Fig.\ \ref{fig:results2d} but for a two-dimensional simulation with
    {\it very}\/ small initial conditions for the
    out-of-plane polarization $A_x$.
    The solid line is the total magnetic energy $\half B^2$;
    the dotted line is the
    energy density $\half B_y^2 + \half B_z^2$ associated with $A_x$.
}
\end{figure}

We can even produce both behaviors in one simulation, shown by the
solid line in Fig.\ \ref{fig:results2dsmallmix}, by initializing the seed
field for
the out-of-plane component $A_x$ much smaller than the other
components.
The solid line shows the total magnetic field energy, which,
for times $m_\infty t \lesssim 80$, behaves similar to the
$A_x=0$ simulation shown by the dashed line of Fig.\
\ref{fig:results2d}: initial exponential growth followed by linear
growth.
The dotted line in Fig.\ \ref{fig:results2dsmallmix} shows
the energy associated with $A_x$, which we take to be
$\half B_y^2 + \half B_z^2 = \half (D_y A_x)^2 + \half (D_z A_x)^2$
in two dimensions.  The importance of $A_x$ continues
to grow throughout. When it eventually catches up to the
other components
at $m_\infty t \simeq 100$, the linear
behavior changes to the continued exponential growth we saw
earlier in simulations where $A_x$ was initialized on the same
footing as the other components.

We have shown simulations that start from tiny initial seed fields for
the instability.  Similar results, shown in the Appendix, apply to
two-dimensional simulations that start from large, non-perturbative
initial configurations: the late-time growth of magnetic energy is
exponential (like one-dimensional simulations) if the out-of-plane polarization
$A_x$ is included in the simulation, and linear (like three-dimensional
simulations) if it is not.

% =========================================================================

\section {Interpretation}
\label {sec:interpret}

\subsection {General}

We can now offer a general characterization of what seems to determine
whether non-abelian effects stop Weibel instability growth in 1D+3V, 2D+3V,
and 3D+3V simulations (at least for the moderately anisotropic
particle distribution we have simulated).
Non-abelian effects will prevent late-time exponential growth if
and only if no gauge-field polarizations are included outside
of the subset of spatial dimensions simulated.
In 1D, we have no choice; unless such polarizations are included,
we cannot simulate magnetic fields and so cannot study the
Weibel instability at all.  In three dimensions, we automatically
include all polarizations.  In two dimensions, we have a choice, with
the results described above.

Formally, the difference between gauge field polarizations in and out
of the subset of spatial dimensions is that those within the subset
(like $A_y$ and $A_z$ in two dimensions) are gauge fields in the
dimensionally reduced theory, while those outside (like $A_x$) are
instead adjoint-charge scalars.  For the two-dimensional theory,
for instance, the Yang-Mills Lagrangian is
\begin {equation}
   \tfrac14 F_{\alpha\beta}^a F^{a \alpha\beta}
   + \tfrac12 (D_\alpha A_x)^a (D^\alpha A_x)^a ,
\end {equation}
where $\alpha$ and $\beta$ are summed over
the 2+1 dimensions $(y,z,t)$.
There is a significant difference between gauge fields and
scalars.  Gauge fields can always be locally transformed away
so that $A=0$ at any particular point, and scalars cannot.
In our two-dimensional theory, the squared amplitude
$A_x^a A_x^a$ of the two-dimensional scalar $A_x$ is locally gauge
invariant and can have physical consequences.

As an example, consider the two-dimensional theory, and imagine that
a component of $A$ has grown very large due to the instability.
Let's focus on distance scales small compared to the wavelength
of the unstable modes.  As a first approximation, treat
$A$ as constant over these small distances and compare and contrast two
situations:
(i) a large,
constant gauge component $A_y$ or $A_z$, and (ii) a large, constant
scalar component $A_x$.
In the first case we can gauge away to $A_y=A_z=0$, and so 
there is no effect at this order of approximation.
One would have to start looking further to the derivatives of A
({\it i.e.} $F^{\mu\nu}$).
In the second case, however, scalar background fields
give rise to the Higgs mechanism.
Except for adjoint color directions that commute with $A_x$,
the gauge fields $A_y$ and $A_z$ will develop large masses of
order $g A_x$.  If $A_x$ is large enough, this could plausibly
suppress all fluctuations of the fields except in commuting color
directions.  If the unsuppressed fields all commute, then non-abelian
interactions cannot prevent instability growth.  This is equivalent
to the ``abelianization'' conjecture of Ref.\ \cite{AL}, which supposed
that, once non-abelian interactions became important, their dynamics
would approximately abelianize the fields and so allow for continued
instability growth.  
What this conjecture did not appreciate was
the difference between scalar fields, such as arise in the dimensionally
reduced theory, and true gauge fields.

To make this difference more concrete, we will now analyze a particular
example of possible relevance to the physics of
Weibel instabilities.

% -------------------------------------------------------------------------

\subsection {An Example: Nielsen-Olesen Instabilities}

Three years ago, before any results had been obtained from
three-dimensional simulations,
Berndt M\"uller \cite{berndt}
suggested that three-dimensional instability growth
would eventually have to stop because, even if the gauge fields
did abelianize, Nielsen-Olesen instabilities \cite{NielsenOlesen}
would eventually
destroy nearly-abelian configurations as the fields continued to
grow.  In this section, we will briefly review these instabilities
and discuss how they are suppressed if some components of gauge
fields are replaced by scalar fields due to dimensional reduction.

For definiteness and simplicity, let us focus on SU(2) gauge theory.
Suppose that there is a large magnetic field whose color lies in an
Abelian U(1) subgroup.
In our application, imagine that this is a large, abelianized magnetic
field that might have formed due to the instability and non-abelian
interactions.
Over small enough distance and time scales (to be discussed later),
we can approximate this magnetic field as constant.
So, for instance,
\begin {equation}
   B^a_i(\x,t) = {\bm B}_0 \, \delta^{a3} .
\label {eq:Bfield}
\end {equation}
Now consider
fluctuations about this background field.
If we ignore the effects of the hard particles represented by
$\delta f(\x,\p,t)$ in our application, but just consider pure
Yang-Mills theory of the soft fields, then this is the problem
studied by Nielsen and Olesen \cite{NielsenOlesen}.
In general, the equation of motion for fluctuations $a$ about a
classical background $A$ is
\begin {equation}
   [ D^2 g^{\mu\nu} - 2i g F^{\mu\nu} ]^{ab} a_\nu^b = 0
\label {eq:background}
\end {equation}
if one works in background gauge ($D_\mu a^\mu = 0$) and if $A$
satisfies the Yang-Mills equations of motion [which is the case
for (\ref{eq:Bfield})].  Here $D = \partial - i gA$ is the
background covariant derivative in adjoint color representation.
For the background (\ref{eq:Bfield}),
this equation can be decomposed into different sectors classified
by (i) the color charge $Q=0$ or $\pm1$ of $a$ under the color
generator $T^3$ and (ii) its spin
projection $\ms=0$ or $\pm1$ along the $\B_0$ axis.
Eq.\ (\ref{eq:background}) becomes%
\footnote{
   Specifically,
   $a^{ai}(x) = \sum_{Q,m_s} a_{Q,m_s}(x)\, \zeta_Q^a \xi_{m_s}^i$,
   where $Q$ and $m_s$ run over $0$ and $\pm 1$;
   $\zeta_{\pm 1} = \xi_{\pm 1} = (1,\pm i, 0)$,
   $\zeta_0 = \xi_0 = (0,0,1)$;
   and $a_{-Q,-m_s} = a_{Q,m_s}^*$.
}
\begin {equation}
   ( D^2 + 2\ms Q g B_0 ) \, a_{Q\ms} = 0
\end {equation}
where $D = \partial - i Q g A$ is an Abelian background derivative
with charge $Qg$.  For $A_0=0$, Fourier transforming $t$ to $\omega$
gives
\begin {equation}
   \bigl[ - ( \grad - i Q g \A )^2 - 2\ms Q g B_0 \bigr] \, a_{Qm}
   = \omega^2 \, a_{Q\ms}.
\label {eq:NO1}
\end {equation}
Mathematically, this is a
relativistic generalization of the Schr\"odinger equation
for a particle with spin in a constant magnetic field.
It has precisely the form of the non-relativistic Schr\"odinger
equation with the non-relativistic energy $E$ replaced by
$\omega^2/2M$.
The $2\ms Qg B_0$ term represents the effect on the
energy of the interaction between the gluon spin and the magnetic field.
The values of $\omega^2$ are determined by (\ref{eq:NO1}) to be
the eigenvalues of the Hamiltonian
\begin {equation}
   H_{Q,m_s} = ( \p - Q g \A )^2 - 2\ms Q g B_0 .
\label {eq:NOH}
\end {equation}
For the three-dimensional case,
one may now follow the discussion of Landau levels in any quantum
mechanics textbook.  The eigenvalues are
\begin {equation}
   \omega^2 = p_\parallel^2 + (n+\half) 2 g B_0 - 2\ms Q g B_0 .
\end {equation}
The first term is from the component $p_\parallel$ of momentum parallel
to the magnetic field $\B_0$, 
the second is the Landau level energy in the plane orthogonal to
$\B_0$, and the last is the
spin interaction with $\B_0$.
The interesting case here is $Q=\ms=\pm1$, for which the lowest Landau orbital
gives
\begin {equation}
  \omega^2 = p_\parallel^2 - g B_0 ,
\end {equation}
which is negative for $p_\parallel \le (g B_0)^{1/2}$.
There are therefore
unstable modes in the relativistic, classical field theory problem.%
\footnote{
  Note that there is nothing unstable in the non-relativistic
  Schr\"odinger problem.
  The instability $\omega^2<0$ in the relativistic,
  classical field theory problem simply corresponds to a negative energy
  state
  in the corresponding non-relativistic Schr\"odinger problem.
}
This is the Nielsen-Olesen instability.
Since it produces growth of gauge field fluctuations $a$ with
non-trivial $T^3$ charge---that is, of colors which do not commute
with the original field $A$---it is an instability which will tend to
destroy abelianization of the gauge fields.

But now let's consider the same Hamiltonian (\ref{eq:NOH}) for the
dimensionally-reduced theory.  For example, $p_x = 0$ in a two
dimensional theory of the $yz$ plane, and so (focusing on the
previously unstable case $Q=\ms=\pm1$):
\begin {equation}
   H_{\pm1,\pm1} = (g A_x)^2 + (p_y \mp g A_y)^2 + (p_z \mp g A_z)^2 - 2gB_0 .
\end {equation}
This Hamiltonian is bounded below by $(g A_x)^2 - 2 g B_0$.
If the $A_x$ field is large enough that
\begin {equation}
   (g A_x)^2 > 2 g B_0 ,
\label {eq:NOcondition}
\end {equation}
then there will be no negative eigenvalue and so no Nielsen-Olesen
instability.
In fact, the $(g A_x)^2$ contribution to the
relation that determines $\omega^2$ is just the Higgs mechanism
due to a background scalar field $A_x$, as discussed earlier.

The one-dimensional case is similar, with now
\begin {equation}
   H_{\pm1,\pm1} = (g A_x)^2 + (g A_y)^2 + (p_z \mp g A_z)^2 - 2gB_0
\end {equation}
bounded below by $(g A_x)^2 + (g A_y)^2 - 2 g B_0$.

We can usefully recast some of our conditions for the application to
Weibel instability growth.  By treating the magnetic field as constant,
we found Nielsen-Olesen instabilities in three dimensions with
typical moment $p_\parallel \sim (g B)^{1/2}$ and growth rates
$\Gamma = (-\omega^2)^{1/2} \sim (g B)^{1/2}$.
The constant magnetic field approximation will be okay if these
scales are large compared to the typical momentum scale $k$ of $B$.
So we need $(g B)^{1/2} \gg k$, which is
\begin {equation}
   B \gg \frac{k^2}{g} \,.
\label {eq:NOcondition2}
\end {equation}
This is simply the condition for the field to have grown past the point
where non-Abelian interactions become important.

For the two-dimensional theory, we can recast condition
(\ref{eq:NOcondition}) for suppressing Nielsen-Olesen instabilities
into a similar form.  Using $B_y = D_z A_x \sim k A_x$, the condition is
\begin {equation}
  B \gtrsim \frac{k^2}{g} \left(\frac{B}{B_y}\right)^2 .
\end {equation}
So, unless $B_y$ is an unusually small component of the magnetic
field (because we set $A_x=0$, for example), then the Nielsen-Olesen
instability will be suppressed in the dimensionally reduced theory
just when it first might have become important, as determined by
(\ref{eq:NOcondition2}).

% =========================================================================

\section {Conclusions}
\label {sec:conclusion}

It is sometimes desirable to simplify a theory, such as through
dimensional reduction, in order to reduce the
cost of numerical simulations.
For studying the fate of non-abelian Weibel instabilities, we have
found qualitatively different behavior in lower-dimensional simulations
when polarizations $A$ of the gauge field are included that are not
gauge fields in the lower-dimensional theory.
This rules out using one-dimensional (1D+3V) simulations to investigate
the fate of non-abelian Weibel instabilities.
However, two-dimensional (2D+3V) simulations can be made to behave similar
to the three-dimensional case if the out-of-plane polarization of
$A$ is discarded.
So far, we have not investigated the case of extremely anisotropic
distributions.

% =========================================================================

\begin{acknowledgments}

We are deeply indebted to Guy D. Moore for providing the original
3-dimensional simulation code that was adapted for this project.
We thank Berndt M\"uller for several conversations over the past
few years related to Nielsen-Olesen instabilities.
This work was supported, in part, by the U.S. Department
of Energy under Grant No.~DE-FG02-97ER41027.

\end{acknowledgments}

% =========================================================================

\appendix

\section {Simulation details}
\label {app:lat}

\subsection {Basics}

We use the same simulation method as Ref.\ \cite{linear1}, discretizing
dependence of $W(\x,\v,t)$ on velocity $\v$ in terms of spherical
harmonics $Y_{lm}(\v)$ for $m \le l \le l_{\rm max}$.
For our initial conditions, we choose a seed magnetic field
configuration and start with zero electric field and hard
particle fluctuation $W$.  For simulations with small initial
conditions,
the initial magnetic field was chosen such that the gauge fields
$A$ have a Gaussian distribution with an exponential fall-off in $\k$
as in Ref.\ \cite{linear1}
Our canonical choice of parameters for two-dimensional simulations
was lattice spacing $a = 0.125/m_\infty$ on a
$256\times256$ lattice of physical
size $L^2 = (256 a)^2 = (32/m_\infty)^2$,
and $l_{\rm max} = 12$ with damping on large $l$ as in Ref.\ \cite{linear2}.

A relatively simple way to simulate the out-of-plane polarization $A_x$
in our 2D+3V simulations is to simply run 3D+3V simulations
on $1\times N_y \times N_z$ lattices.  The links $U_x$ in the short
periodic direction of size $N_x=1$ encode the physics of $A_x(y,z,t)$.
However, we found it simpler to adapt three-dimensional simulations
by instead simulating
$2 \times N_y \times N_z$
lattices with initial conditions that are translation invariant in
the short direction $N_x=2$.  The reason is that the procedure for
updating the $W$ field, as described in Sec. V B and V D of Ref.\ \cite{BMR},
involves alternating updating $W$ on only the even spatial sub-lattice
on one time
step and only the odd spatial sub-lattice the next time step.
For each update of $W$ at a site, the algorithm assumes that $W$ at
neighboring sites is fixed. 
That assumption would fail for a
$1 \times N_y \times N_z$ lattice because every site is its own
neighbor.%
\footnote{
  Ignoring this issue and proceeding with a periodic lattice 
  with $N_x=1$
  introduces artificial numerical
  instabilities unless one modifies the $W$ update algorithm.
  This numerical instability can be understood analytically
  by considering a free $W$ field  ($\partial_t W + \v\cdot\grad W = 0$),
  finding the corresponding discrete-time lattice dispersion
  relation 
  for a given algorithm,
  and checking whether the
  solution for the frequency $\omega$ ever produces a positive
  imaginary part.
}

% ------------------------------------------------------------------------

\subsection{Initial conditions}

\begin{figure}[ht]
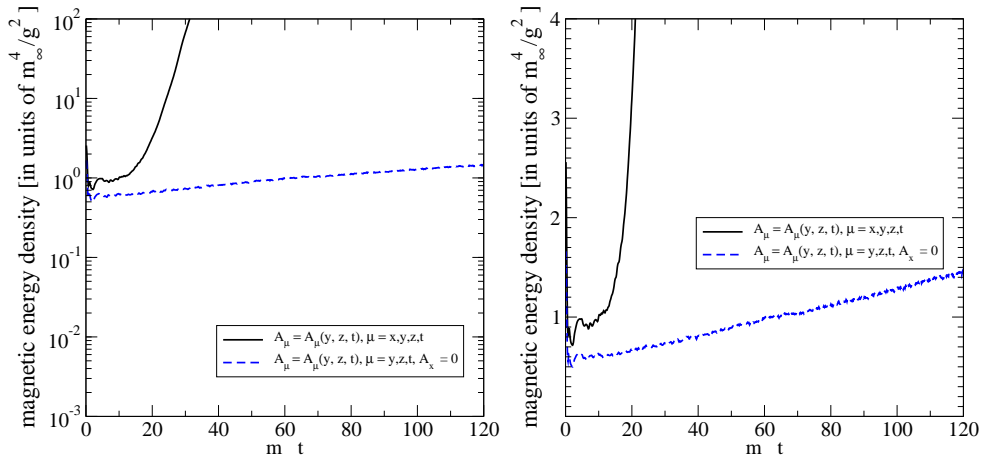

\includegraphics[scale=0.35]{results2dlarge.eps}
\includegraphics[scale=0.35]{results2dlarge_linear.eps}
\caption{%
    \label{fig:results2dlarge}
    As Fig.\ \ref{fig:results2d} but for two-dimensional
    simulations with large initial conditions.
    The initial condition is selected from a thermal
    ensemble with temperature $T=2m_\infty/g^2$, which is then
    gauge-invariantly smeared to remove high-momentum
    components, exactly as in Ref.\ \cite{linear2}.
    [As in Ref.\ \cite{linear2}, the amount of smearing was chosen so
    that perturbatively it would correspond to multiplying the thermal
    spectrum for $A(\k)$ by $\exp(-k^2/4 m_\infty^2)$.]
    For the
    $A_x=0$ simulation, the links $U$ in the $x$ direction
    are then set to unity.
}
\end{figure}

\begin{figure}[ht]
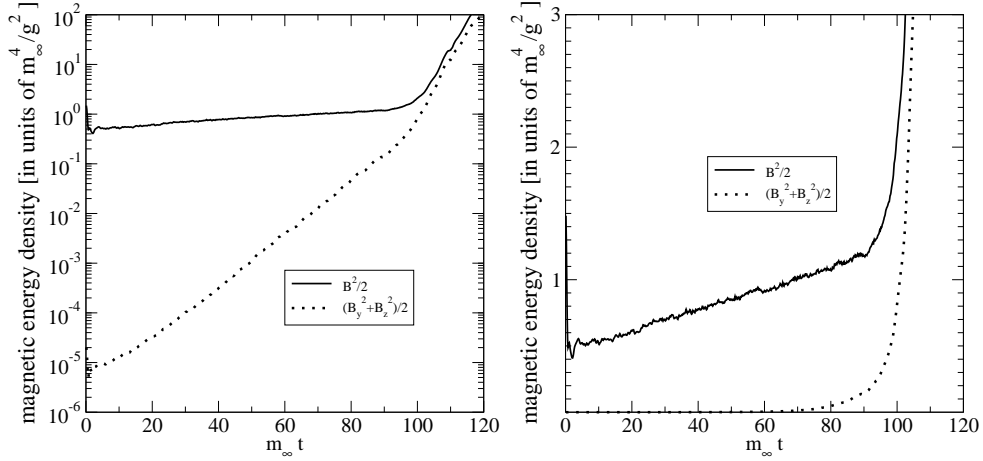

\includegraphics[scale=0.35]{results2dmix.eps}
\includegraphics[scale=0.35]{results2dmix_linear.eps}
\caption{%
    \label{fig:results2dmix}
    As Fig.\ \ref{fig:results2dsmallmix} but the initial conditions
    for the 2-dimensional gauge fields $A_y$ and $A_z$ are large,
    while the out-of-plane component $A_x$ is initialized small.
}
\end{figure}

The simulations discussed in the main text all started with tiny seeds for
the unstable modes.  They all have initial exponential growth, as
predicted by a perturbative analysis of the instability.
In Ref.\ \cite{linear2}, it was noted that linear 
energy growth sets in immediately in three-dimensional simulations
if one instead starts with non-perturbatively large seeds for the
unstable modes.
To check the robustness of our conclusions, we have made similar
simulations in two dimensions.
The results are shown in Fig.\ \ref{fig:results2dlarge} and
\ref{fig:results2dmix}.
As with the case of small initial conditions, the late-time behavior
is exponential if $A_x$ is included and linear if not.

% ------------------------------------------------------------------------

\subsection{Systematic errors}

\begin{figure}[ht]
\includegraphics[scale=0.35]{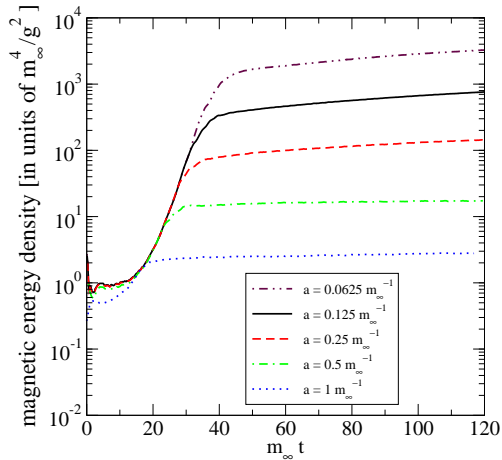}
\caption{%
    \label{fig:Strong_diff_latt_spacing}
    Magnetic energy vs.\ time showing results for several different
    choices of lattice spacing for a volume $L^2 = (32/m_\infty)^2$
    with all polarizations retained (i.e.\ including $A_x$).
}
\end{figure}

\begin{figure}[ht]
\includegraphics[scale=0.35]{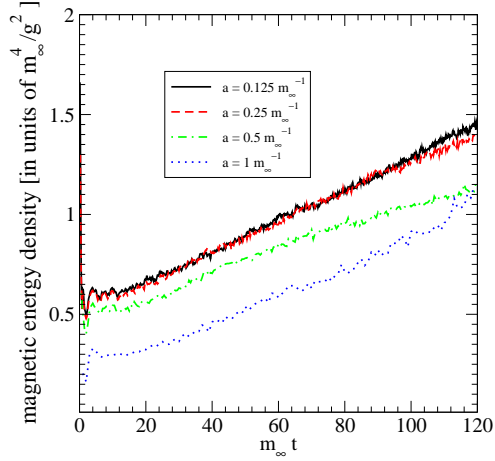}
\caption{%
    \label{fig:Strong_diff_latt_spacing_noAshort}
    Linear plot of
    Magnetic energy vs.\ time showing results for several different
    choices of lattice spacing for a volume $L^2 = (32/m_\infty)^2$
    with out-of-plane polarization $A_x$ not included.
}
\end{figure}

\begin{figure}[ht]
\includegraphics[scale=0.35]{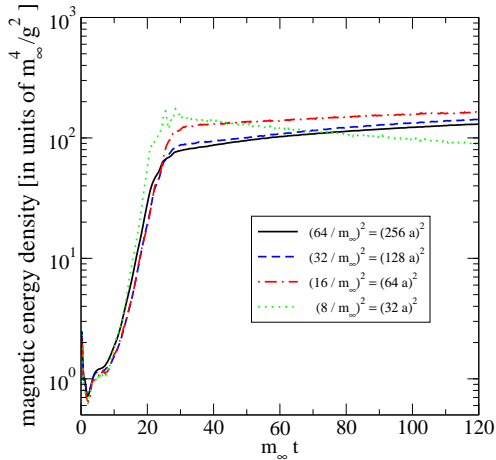}
\caption{%
    \label{fig:Strong_diff_volume}
    Magnetic energy vs.\ time showing results for several different
    physical volumes with lattice spacing $a = 0.25/m_\infty$.
    $A_x$ is included in these simulations.
}
\end{figure}

\begin{figure}[ht]
\includegraphics[scale=0.35]{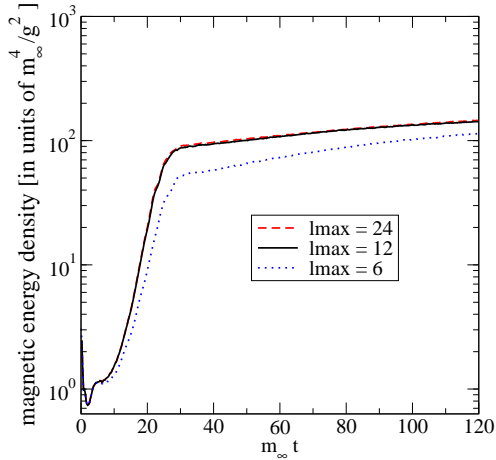}
\caption{%
    \label{fig:Strong_diff_lmax}
    Magnetic energy vs.\ time showing results for several different
    choices of $l_{max}$ with $a = 0.25/m_\infty$ and
    $L^2 = (32/m_\infty)^2$.
    $A_x$ is included in these simulations.
}
\end{figure}

Ref.\ \cite{linear1} provides many checks that their three-dimensional
simulation results are not significantly affected by discretization
effects.
Because the canonical simulation parameters we use for two-dimensional
simulations in this paper are as good or better than the canonical
parameters of Ref.\ \cite{linear1}, one might expect that there are
no significant discretization effects.  To be thorough, however, we
present studies of lattice spacing dependence, volume dependence, and
$l_{\rm max}$ dependence in 
Figs.\ \ref{fig:Strong_diff_latt_spacing}--\ref{fig:Strong_diff_lmax}.
We show here results for simulations with large, non-perturbative
initial conditions, such as discussed above; results are similar for
small initial conditions.

The curves in Fig.\ \ref{fig:Strong_diff_latt_spacing} are like the
solid line in left-hand plot of Fig. \ref{fig:results2dlarge},
where we saw continued exponential growth,
except that here we show explicitly what happens when the fields get
so large that they become limited by the effects of the discrete
lattice.  The proof that this limit is a lattice artifact is that
it increase steadily as the lattice spacing is reduced,
as seen in Fig.\ \ref{fig:Strong_diff_latt_spacing}.
In contrast, the curves in
Fig.\ \ref{fig:Strong_diff_latt_spacing_noAshort} show the case
$A_x=0$.  These curves are like the
dashed line in the right-hand plot of Fig.\ \ref{fig:results2dlarge},
where we saw late-time linear energy growth.
We see here that this behavior is robust as we decrease the lattice
spacing, and that $a = 0.25/m_\infty$ is an adequately small
spacing for our simulations.

Fig.\ \ref{fig:Strong_diff_volume} shows an example of dependence on
physical two-dimensional
volume $L^2$ for the case of continued exponential growth.
(The apparent end to exponential growth shown in this figure is
the same finite-spacing lattice artifact seen in Fig.\
\ref{fig:Strong_diff_latt_spacing}.  We see that nothing changes
qualitatively as the volume is increased: the exponential growth
beyond the non-perturbative scale remains.  In examining the
quantitative variation between the curves, one should keep in mind
that it is not possible to use the same initial conditions for
simulations with different volume: in addition to systematic
effects, there is statistical variation
reflecting the random choice of initial conditions.

Finally, Fig.\ \ref{fig:Strong_diff_lmax} is the analogous figure
for $l_{\rm max}$ dependence.  We see that $l_{\rm max}=12$ is
an adequate approximation to the large $l_{\rm max}$ limit
for our simulations.

%\begin{figure}[ht]
%\includegraphics[scale=0.35]{Strong_diff_seed.eps}
%\caption{%
 %   \label{fig:Strong_diff_seed}
 %   Magnetic energy vs.\ time showing results for several choices of seed.
 %   These simulations have $a = 0.25/m_\inf$, $l_{max} = 12$ and
 %   $L^2 = (32/m_\inf)^2$. 
%}
%\end{figure}

%%%%%%%%%%%%%%%%%%%%%%%%%%%%%%%%%%%%%%%%%%%%%%%%%%%%%%%%%%%%%%%%%%%%%%%%%%%

% =========================================================================

\end {document}